\documentclass{elsart}
\usepackage{graphicx}
\usepackage{amssymb}
\newcommand{\re}{\mathop{\mathrm{Re}}}
\newcommand{\tr}{\mathop{\mathrm{Tr}}}

\begin{document}
\begin{frontmatter}
\title{Time problem in quantum mechanics and weak measurements}
\author{J. Ruseckas\corauthref{cor}}
\corauth[cor]{Corresponding author.}
 \ead{ruseckas@itpa.lt}
 \and
\author{B. Kaulakys}
\address{Institute of Theoretical Physics and Astronomy,\\
A. Go\v{s}tauto 12, 2600 Vilnius, Lithuania}

\begin{abstract}
A question of the time the system spends in the specified state, when the final
state of the system is given, is raised. The model of weak measurements is used
to obtain the expression for the time. The conditions for determination of such
a time are obtained.
\end{abstract}
\end{frontmatter}

\section{Introduction}
Time plays a special role in quantum mechanics. Unlike other observables, time
remains a classical variable. It cannot be simply quantized because, as it is
well known, the self-adjoint operator of time does not exist for bounded
Hamiltonians. The problems with the time rise from the fact that in quantum
mechanics many quantities cannot have definite values simultaneously. The
absence of the time operator makes this problem even more complicated. However,
in practice the time often is important for an experimenter. If quantum
mechanics can correctly describe the outcomes of the experiments, it must also
give the method for the calculation of the time the particle spends in some
region.

The most-known problem of time in quantum mechanics is the so-called ''tunneling
time problem''. There had been many attempts to define physical time for quantum
mechanical tunneling processes, since the question was raised by MacColl
\cite{maccol} in 1932. This question is still the subject of much controversy,
since numerous theories contradict each other in their predictions for ''the
tunneling time'' \cite{haugestovneng,olkhovskyrecami,landauermartin}.  We can
raise another, more general, question about the time. Let us consider a system
which evolves with time. Let $\chi$ is one of the observables of the system.
During the evolution the value of $\chi$ changes. We are considering a subset
$\Gamma$ of possible values of $\chi$. The question is {\em how much time the
values of $\chi$ belong to this subset?}.

There is another version of the question. If we know the final state of the
system, we may ask how much time the values of $\chi$ belong to the subset under
consideration when the system evolves from the initial to the definite final
state. The question about the tunneling time belongs to such class of the
problems. Really, in the tunneling time problem we ask about the duration the
particle spends in a specified region of the space and we know that the particle
has tunneled, i.e., it is on the other side of the barrier. We can expect that
such a question may not always be answered. Here our goal is to obtain the
conditions under which it is possible to answer such a question.

One of the possibilities to solve the problem of time is to answer what exactly
the word ``time'' means. The meaning of every quantity is determined by the
procedure of measurement. Therefore, we have to construct a scheme of an
experiment (this can be a {\em gedanken\/} experiment) to measure the quantity
with the properties corresponding to the classical time.

The experiment for the measurement of time must obey certain conditions. The
time in classical mechanics describes not a single state of the system but the
process of the evolution. This property is an essential concept of the time. We
speak about the time belonging to a certain evolution of the system. If the
measurement of the time disturbs the evolution we cannot attribute this
measured duration to the undisturbed evolution. Therefore, we should require
that the measurement of the time does not disturb the motion of the system.
This means that the interaction of the system with the measuring device must be
weak. In quantum mechanics this means that we cannot use the strong
measurements described by the von-Neumann's projection postulate. We have to
use the weak measurements of Aharonov, Albert and Vaidman
\cite{AAV1,AAV,WM2,AV,AV2,AV3}, instead.

We proceed as follows: In Secs.\ \ref{sec:mod} and \ref{sec:mes}, we present the
model and the procedure of the time measurement, respectively. In Sec.\ 
\ref{sec:condprob} we modify the proposed procedure of the time measurement to
make the distinction between different final states of the system. This
procedure allows us to determine when the concept of the time with the known
final state is correct in quantum mechanics. In Sec.\ \ref{sec:exampl} an
example of application of our formalism for the two-level system is presented.
Section \ref{sec:concl} summarizes our findings.

\section{The model of the time measurement}
\label{sec:mod}

We consider a system evolving with time. One of the quantities describing the
system is $\chi$. Operator $\hat{\chi}$ corresponds to this quantity. For
simplicity we assume that the operator $\hat{\chi}$ has a continuous spectrum.
The case with discrete spectrum will be considered later.

The measuring device interacts with the system only if $\chi$ is near some
point $\chi_{D}$, depending only on the detector. If we want to measure the
time the system is in a large region of $\chi$, we have to use many detectors.
In the case of tunneling a similar model had been introduced by A.~M.~Steinberg
\cite{steinberg} and developed in our paper \cite{JR}. The strong limit of such
a model for analysis of the measurement effect for the quantum jumps has been
used in Ref. \cite{JR2}.

In order the weak measurements can provide the meaningful information, the
measurements have to be performed on an ensemble of identical systems. Each
system with its own detector is prepared in the same initial state. After time
$t$ the readings of the detectors are collected and averaged.

Our model consists of the system {\bf S} under consideration and of the detector
{\bf D}. The Hamiltonian is
\begin{equation}
\hat{H}=\hat{H}_{\mathrm{S}}+\hat{H}_{\mathrm{D}}+\hat{H}_{\mathrm{I}}
\end{equation}
where $\hat{H}_{\mathrm{S}}$ and $\hat{H}_{\mathrm{D}}$ are the Hamiltonians of
the system and of the detector, respectively, and the operator
\begin{equation}
\hat{H}_{\mathrm{I}}=\gamma \hat{q}\hat{D}\left(\chi_{\mathrm{D}}\right)
\label{interact}
\end{equation}
represents the interaction between the system and the detector. The interaction
term (\ref{interact}) only slightly differs from the one used by Aharonov,
Albert and Vaidman \cite{AAV}. The operator $\hat{q}$ acts in the Hilbert space
of the detector. We require a continuous spectrum of the operator $\hat{q}$.
For simplicity, we can consider this operator as the coordinate of the
detector. The operator $\hat{D}\left(\chi_{\mathrm{D}}\right) $ acts in the
Hilbert space of the system. In an ideal case the operator $\hat{D}\left(
\chi_{\mathrm{D}}\right) $ can be expressed as $\delta $ function
\begin{equation}
\hat{D}\left(\chi_{\mathrm{D}}\right)=\left| \chi_{\mathrm{D}}\right\rangle
\left\langle \chi_{\mathrm{D}}\right| =\delta \left(
\hat{\chi}-\chi_{\mathrm{D}}\right) . \label{delta}
\end{equation}
Parameter $\gamma $ in Eq.\ (\ref{interact}) characterizes the strength of the
interaction. A very small parameter $\gamma $ ensures the undisturbance of the
system's evolution.

Hamiltonian (\ref{interact}) with $\hat{D}$ given by (\ref{delta}) represents
the constant force acting on the detector {\bf D} when the quantity $\chi$ is
very close to the value $\chi_{\mathrm{D}}$. This force induces the change of
the detector's momentum. From the classical point of view, the change of the
momentum is proportional to the time the particle spends in the region around
$\chi_{\mathrm{D}}$ and the coefficient of proportionality equals to the force
acting on the detector. We assume that the change of the mean momentum of the
detector is proportional to the time the constant force acts on the detector and
that the time the particle spends in the detector's region is the same as the
time the force acts on the detector.

We can replace the $\delta$ function by the narrow rectangle of the height $1/L$
and of the width $L$ in the $\chi$ space. From Eq.\ (\ref{interact}) it follows
that the force acting on the detector when the particle is in the region around
$\chi_{\mathrm{D}}$ is $F=-\gamma/L$.  The time the particle spends until time
moment $t$ in the unit length region is
\begin{equation}
\tau\left(t\right)=-\frac{1}{\gamma}\left(\left\langle p_{\mathrm{q}}\left(
t\right)\right\rangle -\left\langle p_{\mathrm{q}}\right\rangle \right) 
\label{timedef}
\end{equation}
where $\left\langle p_{\mathrm{q}}\right\rangle$ and $\left\langle
p_{\mathrm{q}}\left(t\right) \right\rangle$ are the mean initial momentum and
momentum after time $t$, respectively.  If we want to find the time the system
spends in the region of the finite width, we have to add many times
(\ref{timedef}).

When the operator $\hat{\chi}$ has a discrete spectrum, we may ask how long the
quantity $\chi$ has the value $\chi_{\mathrm{D}}$. To answer this question the
detector must interact with the system only when $\chi=\chi_{\mathrm{D}}$. In
such a case the operator $\hat{D}\left(\chi_{\mathrm{D}}\right)$ takes the form
\begin{equation}
\hat{D}\left(\chi_{\mathrm{D}}\right)=\left|\chi_{\mathrm{D}}\right\rangle
\left\langle \chi_{\mathrm{D}}\right| .
\end{equation}
The force, acting on the detector in this case equals to $-\gamma$. The time
the quantity $\chi$ has the value $\chi_{\mathrm{D}}$ is given by Eq.\
(\ref{timedef}), too. Further formulae do not depend on the spectrum of the
operator $\hat{\chi}$.

In the time moment $t=0$ the density matrix of the whole system is $\hat{\rho}
\left(0\right)=\hat{\rho}_{\mathrm{S}}\left(0\right)\otimes
\hat{\rho}_{\mathrm{D}}\left(0\right)$, where $\hat{\rho}_{\mathrm{S}}\left(
0\right)$ is the density matrix of the system and $\hat{\rho}_{\mathrm{D}}
\left(0\right)=\left|\Phi\right\rangle\left\langle\Phi\right|$ is the density
matrix of the detector with $\left|\Phi\right\rangle$ being the normalized
vector in the Hilbert space of the detector. After the interaction the density
matrix of the detector is $\hat{\rho}_{\mathrm{D}}\left(t\right)
=\tr_{\mathrm{S}}\left\{\hat{U}\left(t\right)\left(\hat{\rho}_{\mathrm{S}}\left(
0\right)\otimes\left|\Phi\right\rangle\left\langle\Phi\right|\right)
\hat{U}^{\dag}\left( t\right) \right\}$ where $\hat{U}\left( t\right) $ is
the evolution operator.

Later on, for simplicity we will neglect the Hamiltonian of the detector. Then,
the evolution operator approximately equals to the operator $\hat{U}\left(
t,\gamma\hat{q}\right)$ with $\hat{U}\left(t,\alpha\right) $ being the solution
of the equation
\begin{equation}
\mathrm{i}\hbar\frac{\partial}{\partial t}\hat{U}\left( t,\alpha\right) =\left(
\hat{H}_{\mathrm{S}}+\alpha\hat{D}\left(\chi_{\mathrm{D}}\right)\right)
\hat{U}\left( t,\alpha \right) . \label{aproxevol}
\end{equation}
After such assumptions we can obtain the dwell time for our model explicitly.

\section{The dwell time}
\label{sec:mes}

We can expand the operator $\hat{U}\left( t,\gamma \hat{q}\right) $ into the
series of the parameter $\gamma $, assuming that $\gamma $ is small.
Introducing the operator $\hat{D}\left(\chi_{\mathrm{D}}\right)$ in the
interaction representation
\begin{equation}
\tilde{D}\left(\chi_{\mathrm{D}},t\right)=\hat{U}^{\dag}_{\mathrm{S}}\left(
t\right)\hat{D}\left(\chi_{\mathrm{D}}\right)\hat{U}_{\mathrm{S}}\left(
t\right)
\end{equation}
where $\hat{U}_{\mathrm{S}}\left(t\right)$ is the evolution operator of
unperturbed system we obtain the first-order approximation for the operator
$\hat{U}\left( t,\gamma \hat{q}\right)$,
\begin{equation}
\hat{U}\left(t,\gamma\hat{q}\right)\approx\hat{U}_{\mathrm{S}}\left(t\right)
\left(1+\frac{\gamma\hat{q}}{\mathrm{i}\hbar}\int_{0}^{t}\d t_{1}
\tilde{D}\left(\chi_{\mathrm{D}},t_{1}\right)\right) .
\end{equation}
For shortening of the notation we introduce the operator
\begin{equation}
\hat{F}\left(\chi_{\mathrm{D}},t\right)=\int_{0}^{t}\d t_{1}
\tilde{D}\left(\chi_{\mathrm{D}},t_{1}\right). \label{opf}
\end{equation}

From the density matrix of the detector after the measurement in the first-order
approximation we find that the average change of the momentum of the detector
during the time $t$ is
$-\gamma\left\langle\hat{F}\left(\chi_{\mathrm{D}},t\right)\right\rangle$. From
Eq.\ (\ref{timedef}) we obtain the dwell time until time moment $t$,
\begin{equation}
\tau\left(\chi,t\right)=\left\langle\hat{F}\left(\chi,t\right)\right\rangle .
\label{fulltime}
\end{equation}
The time spent in the region $\Gamma$ is
\begin{equation}
t\left(\Gamma;t\right)=\int_{\Gamma}\d \chi\tau\left(\chi,t\right)= \int_0^t\d
t'\int_{\Gamma}\d \chi P\left(\chi,t'\right), \label{eq:dw}
\end{equation}
where $P\left(\chi,t'\right)=\left\langle\tilde{D}\left(\chi,t\right)
\right\rangle$ is the probability for the system to have the value $\chi$ at
time moment $t'$.

In the case when $\chi$ is the coordinate of the particle Eq.\ (\ref{eq:dw})
yields the well-known expression for the dwell time \cite{olkhovskyrecami,JR}.
This time is the average over the entire ensemble of the systems, regardless of
their final states.

\section{The time on condition that the system is in the given final
state}\label{sec:condprob}

Further we will consider the case when the final state of the system is known.
We ask {\em how much time the values of $\chi$ belong to the subset under
  consideration, $\Gamma$, on condition that the system evolves to the
  definite final state $f$}. More generally, we might know that the final state
of the system belongs to the certain subspace ${\mathcal H}_{\mathrm{f}}$ of
system's Hilbert space.

The projection operator which projects the vectors from the Hilbert space of
the system into the subspace ${\mathcal H}_{\mathrm{f}}$ of the final states is
$\hat{P}_{\mathrm{f}}$. As far as our model gives the correct result for the
time averaged over the entire ensemble of the systems, we can try to take the
average only over the subensemble of the systems with the given final states.
We measure the momenta $p_{\mathrm{q}}$ of each measuring device after the
interaction with the system. Subsequently we perform the final, postselection
measurement on the systems of our ensemble. Then we collect the outcomes
$p_{q}$ only of the systems the final state of which turns out to belong to the
subspace ${\mathcal H}_{\mathrm{f}}$.

The joint probability that the state of the system belongs to ${\mathcal
H}_{\mathrm{f}}$ {\em and\/} the detector has the momentum $p_{\mathrm{q}}(t)$
at the time moment $t$ is $W\left( P_{\mathrm{f}},p_{\mathrm{q}};t\right)
=\tr\left\{\hat{P}_{\mathrm{f}}\left|p_{\mathrm{q}}\right\rangle \left\langle
p_{\mathrm{q}}\right|\hat{\rho}\left(t\right)\right\}$, where $\left|
p_{\mathrm{q}}\right\rangle $ is the eigenfunction of the momentum operator
$\hat{p}_{\mathrm{q}}$. In quantum mechanics the probability that two
quantities simultaneously have definite values does not always exist. If the
joint probability does not exist then the concept of the conditional
probability is meaningless. However, in our case operators
$\hat{P}_{\mathrm{f}}$ and $\left| p_{\mathrm{q}}\right\rangle \left\langle
p_{\mathrm{q}}\right| $ act in different spaces and commute, therefore, the
probability $W\left( P_{\mathrm{f}},p_{\mathrm{q}};t\right) $ exists.

Let us define the conditional probability, i.e., the probability that the
momentum of the detector is $p_{\mathrm{q}}$ provided that the state of the
system belongs to ${\mathcal H}_{\mathrm{f}}$. This probability is given
according to the Bayes's theorem
\begin{equation}
W\left( p_{\mathrm{q}};t\left|P_{\mathrm{f}}\right.\right)=\frac{W\left(
P_{\mathrm{f}},p_{\mathrm{q}};t\right)}{W\left(P_{\mathrm{f}};t\right) }
\label{condprob}
\end{equation}
where $W\left(P_{\mathrm{f}};t\right) =\tr\left\{\hat{P}_{\mathrm{f}}\hat{\rho%
}\left( t\right) \right\} $ is the probability that the state of the system
belongs to the subspace ${\mathcal H}_{\mathrm{f}}$. The average momentum of
the detector on condition that the state of the system belongs to the
subspace ${\mathcal H}_{\mathrm{f}}$ is
\begin{equation}
\left\langle p_{\mathrm{q}}\left(t\right)\right\rangle=\int p_{\mathrm{q}}\d
p_{\mathrm{q}}W\left(p_{\mathrm{q}};t\left| P_{\mathrm{f}}\right. \right) .
\label{condave}
\end{equation}

From Eqs.\ (\ref{timedef}) and (\ref{condave}), in the first-order approximation
we obtain the duration on condition that the final state of the system
belongs to the subspace ${\mathcal H}_{\mathrm{f}}$
\begin{eqnarray}
\tau_{\mathrm{f}}\left( \chi,t\right) &=& \frac{1}{2\left\langle
\tilde{P}_{\mathrm{f}}\left(t\right)\right\rangle}\left\langle
\tilde{P}_{\mathrm{f}}\left(t\right)\hat{F}\left(\chi,t\right)+\hat{F}\left(
\chi,t\right)\tilde{P}_{\mathrm{f}}\left( t\right)
\right\rangle  \nonumber \\
&& +\frac{1}{\mathrm{i}\hbar \left\langle \tilde{P}_{\mathrm{f}}\left(t\right)
\right\rangle }\left( \left\langle q\right\rangle \left\langle
p_{\mathrm{q}}\right\rangle-\re\left\langle
\hat{q}\hat{p}_{\mathrm{q}}\right\rangle \right) \left\langle \left[
\tilde{P}_{\mathrm{f}}\left( t\right) ,\hat{F}\left( \chi,t\right) \right]
\right\rangle . \label{condtime}
\end{eqnarray}

Eq.\ (\ref{condtime}) consists of two terms and we can introduce two
expressions with the dimension of time
\begin{eqnarray}
\tau_{\mathrm{f}}^{(1)}\left(\chi,t\right) &=&\frac{1}{2\left\langle
\tilde{P}_{\mathrm{f}}\left(t\right)\right\rangle}\left\langle
\tilde{P}_{\mathrm{f}}\left(t\right)\hat{F}\left(\chi,t\right)+\hat{F}\left(
\chi,t\right)
\tilde{P}_{\mathrm{f}}\left(t\right)\right\rangle ,  \label{timere} \\
\tau_{\mathrm{f}}^{(2)}\left(\chi,t\right) &=&\frac{1}{2\mathrm{i}\left\langle
\tilde{P}_{\mathrm{f}}\left(t\right)\right\rangle}\left\langle \left[
\tilde{P}_{\mathrm{f}}\left(t\right),\hat{F}\left(\chi,t\right) \right]
\right\rangle . \label{timeim}
\end{eqnarray}
Then the time the system spends in the subset $\Gamma$ on condition that
the final state of the system belongs to the subspace ${\mathcal
  H}_{\mathrm{f}}$ can be rewritten in the form
\begin{equation}
\tau_{\mathrm{f}}\left(\chi,t\right)=\tau_{\mathrm{f}}^{(1)}\left(\chi,t\right)+
\frac{2}{\hbar}\left(\left\langle q\right\rangle\left\langle p_{\mathrm{q}}
\right\rangle-\re\left\langle\hat{q}\hat{p}_{\mathrm{q}}\right\rangle\right)
\tau_{\mathrm{f}}^{(2)}\left(\chi,t\right). \label{condtime2}
\end{equation}
The quantities $\tau_{\mathrm{f}}^{(1)}(\chi,t)$ and
$\tau_{\mathrm{f}}^{(2)}(\chi,t)$ are related to the real and imaginary parts of
the complex time, introduced by D.\ Sokolovski {\it et.\ al\/}
\cite{sokolovskibaskin}. In our model the quantity $\tau_{\mathrm{f}}(\chi,t)$
is real, contrary to the complex-time approach. The components of time
$\tau_{\mathrm{f}}^{(1)}$ and $\tau_{\mathrm{f}}^{(2)}$ are real, too.
Therefore, this time can be interpreted as the duration of an event.

If the commutator $\left[\tilde{P}_{\mathrm{f}}\left( t\right) ,\hat{F}\left(
    \chi,t \right) \right] $ in Eq.\ (\ref{condtime}) is not zero then, even in
the limit of the very weak measurement, the measured value depends on the
particular detector. This fact means that in such a case we cannot obtain the
{\em definite\/} value for the conditional time. Moreover, the coefficient
$(\langle q\rangle\langle p_{\mathrm{q}}\rangle-\re\langle\hat{q}
\hat{p}_{\mathrm{q}}\rangle)$ may be zero for the specific initial state of the
detector, e.g., for the Gaussian distribution of the coordinate $q$ and momentum
$p_{\mathrm{q}}$.

The conditions of the possibility to determine the time uniquely in a
case when the final state of the system is known takes the form
\begin{equation}
\left[\tilde{P}_{\mathrm{f}}\left(t\right),\hat{F}\left(\chi,t\right)\right]=0
. \label{eq:poscond}
\end{equation}
This result can be understood basing on general principles of quantum mechanics,
too. We ask {\em how much time the values of $\chi$ belong to the certain subset
  when the system evolves to the given final state}. We know with certainty the
final state of the system. In addition, we want to have some information about
the values of the quantity $\chi$. However, if we know the final state with
certainty, we may not know the values of $\chi$ in the past and, vice versa, if
we know something about $\chi$, we may not definitely determine the final state.
Therefore, in such a case the question about the time when the system evolves to
the given final state cannot be answered definitely and the conditional time has
no reasonable meaning.

The quantity $\tau_{\mathrm{f}}\left(t\right)$ according to Eqs.\
(\ref{condtime}) and (\ref{timere}) has many properties of the classical time.
So, if the final states $\left\{f\right\}$ constitute the full set, then the
corresponding projection operators obey the equality of completeness $\sum_f
\hat{P}_{\mathrm{f}}=1$. Then from Eq.\ (\ref{condtime}) we obtain the expression
\begin{equation}
\sum_f \left\langle \tilde{P}_{\mathrm{f}}\left( t\right) \right\rangle
\tau_{\mathrm{f}}\left( \chi,t\right) =\tau\left( \chi,t\right) .
\label{clasprop}
\end{equation}
The quantity $\left\langle \tilde{P}_{\mathrm{f}}\left( t\right) \right\rangle$
is the probability that the system at the time $t$ is in the state $f$. Eq.\
(\ref{clasprop}) shows that the full duration equals to the average over all
possible final states, as it is a case in the classical physics. From Eq.\
(\ref{clasprop}) and Eqs.\ (\ref{timere}), (\ref{timeim}) it follows
\begin{eqnarray}
\sum_f \left\langle \tilde{P}_{\mathrm{f}}\left( t\right) \right\rangle
\tau_{\mathrm{f}}^{(1)}\left( \chi,t\right) &=& \tau\left( \chi,t\right) ,\\
\sum_f \left\langle \tilde{P}_{\mathrm{f}}\left( t\right) \right\rangle
\tau_{\mathrm{f}}^{(2)}\left( \chi,t\right) &=& 0 .
\end{eqnarray}
We suppose that quantities $\tau_{\mathrm{f}}^{(1)}\left(\chi,t\right)$ and
$\tau_{\mathrm{f}}^{(2)}\left(\chi,t\right)$ can be useful even in the case
when the time has no definite value, since in the tunneling time problem the
quantities (\ref{timere}) and (\ref{timeim}) correspond to the real and
imaginary parts of the complex time, respectively \cite{JR}.

The eigenfunctions of the operator $\hat{\chi}$ constitute the full set
$\int\left|\chi\right\rangle\left\langle\chi\right|\d \chi=1$ where the integral
must be replaced by the sum for the discrete spectrum of the operator
$\hat{\chi}$. From Eqs.\ (\ref{delta}), (\ref{opf}), (\ref{condtime}) we obtain
the equality
\begin{equation}
\int \tau_{\mathrm{f}}\left(\chi,t\right)\d \chi=t . \label{eq:eqx}
\end{equation}
Eq.\ (\ref{eq:eqx}) shows that the time during which the quantity $\chi$
has any value equals to $t$, as it is in the classical physics.

\section{Example}
\label{sec:exampl}

The obtained formalism can be applied to the tunneling time problem \cite{JR}.
In this paper, however, we will consider a simpler system than the tunneling
particle, i.e., a two-level system. The system is forced by the perturbation
$V$ which causes the jumps from one state to another. We will determine the
time the system is in the given state.

The Hamiltonian of this system is
\begin{equation}
\hat{H}=\hat{H}_0+\hat{V} 
\end{equation}
where $\hat{H}_0=\hbar\omega\hat{\sigma}_3/2$ is the Hamiltonian of the
unperturbed system and $\hat{V}=v\hat{\sigma}_{+}+v^{*}\hat{\sigma}_{-}$ is the
perturbation.  Here $\sigma_1,\sigma_2,\sigma_3$ are Pauli matrices and
$\sigma_{\pm}=\frac{1}{2}\left( \sigma_1 \pm i\sigma_2\right)$. The Hamiltonian
$\hat{H}_0$ has two eigenfunctions $\left|0\right\rangle$ and
$\left|1\right\rangle$ with the eigenvalues $-\hbar\omega/2$ and
$\hbar\omega/2$, respectively. The initial state of the system is
$\left|0\right\rangle$.

From Eq.\ (\ref{fulltime}) we obtain the times the system spends in the energy
levels $0$ and $1$, respectively,
\begin{eqnarray}
\tau\left(0,t\right) &=&\frac{1}{2}\left(1+\frac{\omega^2}{\Omega^2}\right)
t+\frac{1}{2\Omega}\sin\left(\Omega t\right) \left(1-
\frac{\omega^2}{\Omega^2}\right) , \label{eq:31}\\
\tau\left(1,t\right)&=&\frac{1}{2}\left(1- \frac{\omega^2}{\Omega^2}\right)
t-\frac{1}{2\Omega}\sin\left( \Omega t\right)
\left(1-\frac{\omega^2}{\Omega^2}\right) \label{eq:32}
\end{eqnarray}
where $\Omega=\sqrt{\omega^{2}+4\frac{\left|v\right|^2}{\hbar^2}}$.  From Eqs.
(\ref{timere}) and (\ref{timeim}) we can obtain the conditional time. The
components $\tau^{(1)}$ (\ref{timere}) and $\tau^{(2)}$ (\ref{timeim}) of the
time the system spends in the level $0$ on condition that the final state
is $\left|1\right\rangle$ are
\begin{eqnarray}
\tau_1^{(1)}\left( 0,t\right) &=& \frac{t}{2},\label{eq:34}\\
\tau_1^{(2)}\left( 0,t\right) &=& \frac{\omega}{2\Omega}\left( 1-t\cot \left(
\frac{\Omega}{2} t\right) \right) .
\end{eqnarray}
When $\Omega t=2\pi n$, $n\in \Zset$, the quantity $\tau_1^{(2)}\left(
  0,t\right)$ tends to infinity. This is because at these time moments the
system is in state $|1\rangle$ with the probability $0$ and we cannot consider
the interaction with the detector as very weak.

\begin{figure}
\begin{center}
\includegraphics*[width=.6\hsize]{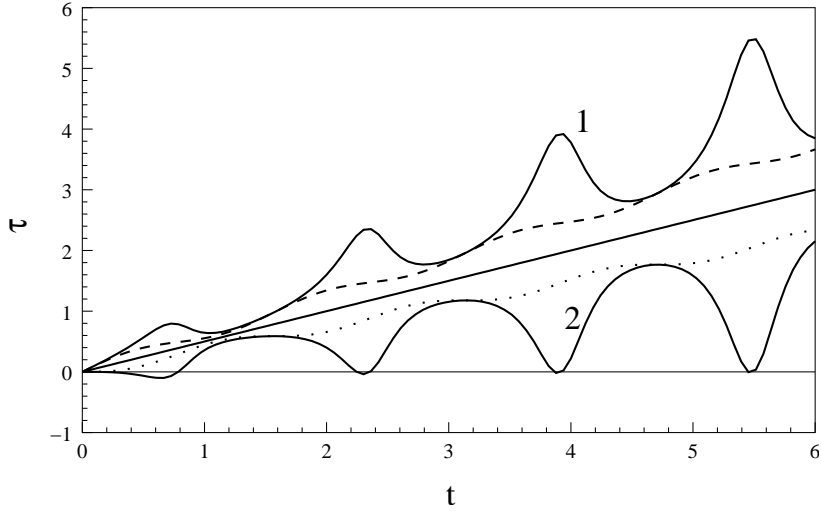}
\end{center}
\caption{The times the system spends in the energy levels $0$, $\tau\left(
    0,t\right)$ (dashed line) and level $1$, $\tau\left( 1,t\right)$ (dotted
  line), according to Eqs.\ (\ref{eq:31}) and (\ref{eq:32}), respectively. The
  quantity $\tau_1^{(1)}\left( 0,t\right) $, Eq.\ (\ref{eq:34}) is shown as
  solid straight line. The quantities $\tau^{(1)}_0(0,t)$ (1) and
  $\tau^{(1)}_0(1,t)$ (2) are calculated according to Eqs.\ (\ref{eq:28}) and
  (\ref{eq:30}), respectively. The parameters are $\omega=2$ and $\Omega=4$.}
\label{dwt}
\end{figure}
The components of the time (\ref{timere}) and (\ref{timeim}) the system spends
in level $0$ on condition that the final state is $\left|0\right\rangle$ are
\begin{eqnarray}
\tau_0^{(1)}\left(0,t\right)&=&\frac{\left(1+3\frac{\omega^2}{\Omega^2}\right)
t+\left(1-\frac{\omega^2}{\Omega^2}\right)\left(\frac{2}{\Omega}\sin\left(
\Omega t\right)+t\cos\left(\Omega t\right)\right)}{2\left( \left(
1+\frac{\omega^2}{\Omega^2}\right)+\left(1-\frac{\omega^2}{\Omega^2}\right)
\cos\left(\Omega t\right)\right)} ,\label{eq:28} \\
\tau_0^{(2)}\left(0,t\right)&=&\frac{\frac{\omega}{\Omega}\left(1-\frac{\omega
^2}{\Omega^2}\right)\sin\left(\frac{\Omega}{2}t\right)\left(t\cos\left(
\frac{\Omega}{2}t\right)-\frac{2}{\Omega}\sin\left(\frac{\Omega}{2}t\right)
\right)}{2\left( \left( 1+ \frac{\omega^2}{\Omega^2}\right)+
\left(1-\frac{\omega^2}{\Omega^2}\right) \cos\left(\Omega t\right)\right)} .
\label{eq:29}
\end{eqnarray}

\begin{figure}
\begin{center}
\includegraphics*[width=.4\hsize]{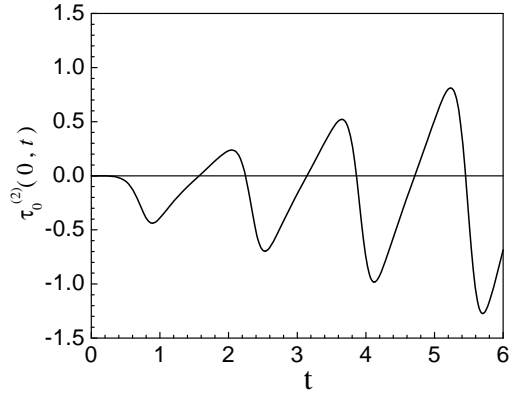}
\end{center}
\caption{The quantity $\tau_0^{(2)}\left( 0,t\right)$, Eq.\ (\ref{eq:29}).
  The parameters are the same as in Fig.\ \protect\ref{dwt}.}
\label{ti00}
\end{figure}
The time the system spends in level $1$ on condition that the final state is
$\left|0\right\rangle$ may be expressed as
\begin{equation}
\tau_0^{(1)}\left( 1,t\right) = \frac{\left( 1- \frac{\omega ^2}{\Omega
^2}\right) \left( t+t\cos\left(\Omega t\right)- \frac{2}{\Omega}\sin\left(
\Omega t\right) \right)}{2\left(\left(1+\frac{\omega^2}{\Omega^2}\right)
+\left(1-\frac{\omega^2}{\Omega^2}\right) \cos\left(\Omega t\right)\right)} .
\label{eq:30}
\end{equation}

The quantities $\tau\left(0,t\right)$, $\tau\left(1,t\right)$,
$\tau_1^{(1)}\left( 0,t\right)$, $\tau_0^{(1)}\left( 0,t\right)$ and
$\tau_0^{(1)}\left(1,t \right)$ are shown in Fig.\ \ref{dwt}. The quantity
$\tau_0^{(2)}\left( 0,t\right)$ is shown in Fig.\ \ref{ti00}.  Note that the
duration with the given final state is not necessarily monotonic as it is with
the full duration, because this final state at different time moments can be
reached by different ways. We can interpret the quantity $\tau_0^{(1)}\left(
  0,t\right)$ as the time the system spends in the level $0$ on condition that
the final state is $\left|0\right\rangle$, but at certain time moments this
quantity is greater than $t$. The quantity $\tau_0^{(1)}\left( 1,t\right)$
becomes negative at certain time moments. This is the consequence of the fact
that for the system under consideration the condition (\ref{eq:poscond}) is not
fulfilled. The peculiarities of the behavior of the conditional times show that
it is impossible to decompose the unconditional time into two components having
all classical properties of the time.

\section{Conclusion}
\label{sec:concl}

We consider the problem of the time in quantum mechanics. The tunneling time
problem is a part of this more general problem. The problem of time is solved
adapting the weak measurement theory to the measurement of time. In this model
expression (\ref{fulltime}) for the duration when the arbitray observable $\chi$
has the certain value is obtained. This result is in agreement with the known
results for the dwell time in the tunneling time problem.

Further we consider the problem of the duration when the observable $\chi$ has
the certain value on condition that the system is in the given final state.
Our model of measurement allows us to obtain the expression (\ref{condtime}) of
this duration as well.  This expression has many properties of the corresponding
classical time.  However, such a duration not always has the reasonable meaning.
It is possible to obtain the duration the quantity $\chi$ has the certain value
on condition that the system is in a given final state only when the condition
(\ref{eq:poscond}) is fulfilled. In the opposite case, there is a dependence in
the outcome of the measurements on particular detector even in an ideal case
and, therefore, it is impossible to obtain the definite value of the duration.
When the condition (\ref{eq:poscond}) is not fulfilled, we introduce two
quantities (\ref{timere}) and (\ref{timeim}), characterizing the conditional
time. These quantities are useful in the case of tunneling and we suppose that
they can be useful also for other problems.


\begin{thebibliography}{00}
\bibitem{maccol}  L. A. MacColl, Phys. Rev. 40 (1932) 621.
\bibitem{haugestovneng}  E. H. Hauge and J. A. St\o vneng, Rev. Mod. Phys.
61 (1989) 917.
\bibitem{olkhovskyrecami}  V. S. Olkhovsky, and E. Recami, Phys. Rep.
214 (1992) 339.
\bibitem{landauermartin}  R. Landauer and Th. Martin, Rev. Mod. Phys.
66 (1994) 217.
\bibitem{AAV1} Y. Aharonov, D. Albert, A. Casher, L. Vaidman, Phys. Lett. A
124 (1987) 199.
\bibitem{AAV} Y. Aharonov, D. Z. Albert, L. Vaidman, Phys. Rev. Lett.
60 (1988) 1351.
\bibitem{WM2} I. M. Duck, P. M. Stevenson, E. C. G. Sudarshan, Phys. Rev. D
40 (1989) 2112.
\bibitem{AV} Y. Aharonov, L. Vaidman, Phys. Rev. A 41 (1990) 11.
\bibitem{AV2} Y. Aharonov, L. Vaidman, J. Phys. A 24 (1991) 2315.
\bibitem{AV3} Y. Aharonov, L. Vaidman, Phys. Scr. T 76 (1998) 85.
\bibitem{steinberg}  A. M. Steinberg, Phys. Rev. A 52 (1995) 32.
\bibitem{JR} J. Ruseckas, Phys. Rev. A 63 (2001) 052107; quant-ph/0101136.
\bibitem{JR2} J.Ruseckas and B. Kaulakys, Phys. Rev. A 63 (2001) 062103.
\bibitem{sokolovskibaskin} D. Sokolovski and L. M. Baskin, Phys. Rev. A
36 (1987) 4604.
\end{thebibliography}
\end{document}